 \documentclass[11pt]{article}
 \usepackage[top=1in,bottom=1in,left=1.5in,right=1.5in]{geometry}
  \usepackage{amsmath,amssymb}
  \usepackage{latexsym}
  \usepackage[dvips]{pstricks} 
  \usepackage[dvips]{graphicx}

  \newcommand{\C}{\mathbb{C}}
  \newcommand{\N}{\mathbb{N}}

  \newcommand{\rH}{\rm{H}}
  
  \newcommand{\lan}{\langle}
  \newcommand{\ran}{\rangle}
  \newcommand{\an}[1]{\lan#1\ran}
  \newcommand{\hs}{\hspace*{\parindent}}
  \newcommand{\proof}{\hs \textbf{Proof.\ }}
  \newcommand{\tr}{\mathop{\mathrm{Tr}}\nolimits}
  \newcommand{\diag}{\mathrm{diag}}

  \newcommand{\qed}{\hspace*{\fill} $\Box$\\}

  \newtheorem{theo}{\bfseries \hs Theorem}[section]

  \newtheorem{lemma}[theo]{\bfseries \hs Lemma}
  \newtheorem{corol}[theo]{\bfseries \hs Corollary}

  \numberwithin{equation}{section} 

\begin{document}

 \title{Remarks on BMV conjecture }
  \author{Shmuel Friedland
 \thanks{Department of Mathematics, Statistics, and Computer Science,
  University of Illinois at Chicago
  Chicago, Illinois 60607-7045, USA, \texttt{E-mail: friedlan@uic.edu}}
  \thanks{Visiting Professor,
  Berlin Mathematical School,
  Institut f\"{u}r Mathematik,
  Technische Universit\"{a}t Berlin,
  Strasse des 17. Juni 136,
  D-10623 Berlin, Germany}}

 \date{April 24, 2008}

 \maketitle

 \begin{abstract}
 We show that for fixed $A,B$, hermitian nonnegative definite
 matrices, and fixed $k$ the coefficients of the $t^k$ in the
 polynomial $\tr (A+tB)^m$ is positive if $\tr AB >0$ and $m>N(A,B,k)$.
 \\
 \noindent
 Keywords and phrases:  Bessis-Moussa-Villani conjecture,
 asymptotic results.

 \end{abstract}

 \section{Introduction}
 It was shown by Lieb-Seiringer \cite{LS} that the Bessis-Moussa-Villani
 conjecture \cite{BMV} to the following statement.  Assume that
 $A,B$ $n\times n$ are hermitian nonnegative definite matrices.
 Then the polynomial $\tr (A+tB)^m$ has nonnegative Taylor coefficients.
 Note that $\tr AB\ge 0$ and $\tr AB=0$ if and only if $AB=0$.
 The aim of this paper to show the following asymptotic result.
 For a fixed integer $k\ge 0$, the coefficients of the $t^k$ in the
 polynomial $\tr (A+tB)^m$ is positive if $\tr AB >0$ and $m>N(A,B,k)$.
 We obtained this result in summer 2007.  Since then another
 proof appeared in \cite{Fle}.
 We also discuss the case of nonnegativity of all coefficients of $t^3$  in $\tr(A+tB)^m$
 for  $n=3$ and all $m$.  I would like to thank to J. Borcea
 for drawing my attention to \cite{Fle}.

 \section{Preliminary results}

 Let $\C^{m\times n}, \rH_n$ be the set of $m\times n$ and $n\times
 n $ hermitian matrices respectively.  For $A\in\C^{n\times n}$ let
 $\tr A$ be the trace of $A$.  Let $A_1,\ldots, A_k\in \C^{n\times
 n}$.  Then $\tr A_1\ldots A_k$ is a cyclic invariant:
 $$\tr A_1A_2\ldots A_{k-1} A_k=\tr A_2A_3\ldots A_k A_1=\ldots=\tr A_k
 A_1\ldots A_{k-1}.$$
 Denote
 \begin{equation}\label{defhurp}
 (A+t B)^m=\sum_{i=0}^m t^i S_{m-i,i}(A,B), \quad A,B\in \C^{n\times n}.
 \end{equation}
 Since
 $$(A+tB)^{m+1}=(A+tB)(A+tB)^m=(A+tB)^m(A+tB)$$
 we deduce
 \begin{eqnarray}\label{basid}
 S_{m+1,k}(A,B)=AS_{m,k}(A,B)+BS_{m+1,k-1}(A,B)=\\
 S_{m,k}(A,B)A +
 S_{m+1,k-1}(A,B)B, \quad  m,k=0,1,\ldots, \nonumber
 \end{eqnarray}
 where we use the convention
 \begin{equation}\label{conven}
 S_{p,q}(A,B)=0 \textrm{ if } min(p,q)<0.
 \end{equation}

 Note that if $A,B\in \rH_n$ are hermitian then
 $S_{p,q}(A,B)\in \rH_n$ for each $p,q\ge 0$.
 For $A\in\rH_n$ we denote by $A\succ 0, A\succeq 0$ the positive
 and the nonnegative definite matrices respectively.
 $\rH_{n,+}:=\{A\in \rH_n:\; A \succeq 0\}$.
 The BMV conjecture is equivalent to $Tr S_{p,q}\ge 0$ for each  $A,B\in \rH_{n,+},p,q\ge 0$.
 \cite{LS}.
 Recall that $\tr AB\ge 0$ if $A,B\succeq 0$.  It is easy to show
 that $\tr S_{p,q}(A,B)\ne 0$ if $\min(p,q)\le 2$.
 (One uses the cyclic invariance of the trace and the fact $\tr CD\ge 0$ if $C,D\succeq 0$).
 By replacing the hermitian pair $(A,B)$ with $(UAU^*,U BU^*)$,
 where  $U$ is a unitary matrix, without
 loss of generality we may assume that
 \begin{equation}\label{Adiag} A=\diag(a_1,\ldots,a_n), \quad
 a_1\ge\ldots\ge a_n\ge 0.
 \end{equation}

 \section{Asymptotic results}

 In view of the results of the results \cite{Hil} it is of interest
 to consider the asymptotic behavior of $Tr S_{m,k}(A,B)$ for a
 fixed  $A,B\in \rH_{n,+},k$ and $m\to\infty$.

 \begin{theo}\label{fasymin}  Let $A,B\in \rH_{n,+}$.  If $Tr AB=0$
 then $AB=0$, Hence $Tr S_{m,k}(A,B)=0$ for any $m,k\ge 1$.
 Assume that $Tr AB>0$.  Let $A$ be a diagonal matrix of the form
 (\ref{Adiag}) and $B=[b_{ij}]_{i,j=1}^n$.
 Let $p\in \an{n}$ be the smallest $j$ such that $a_j b_{jj}>0$.
 Then for each $\epsilon>0, k\in \N$ there exists $N(A,B,\epsilon,k)$ such that
 \begin{equation}\label{asympineq1}
 Tr S_{m,k}(A,B)\ge (1-\epsilon)b_{pp}^k a_p^m {m+k \choose k} \textrm{ for }
 m> N(A,B,\epsilon,k).
 \end{equation}
 Furthermore, assume that $a_p=\ldots=a_{p+l-1}>a_{p+l}, l\in [1,n]$.
 Denote by $C=[b_{ij}]_{i,j=p}^{p+l-1}\in \rH_{l,+}$.
 Then
 \begin{equation}\label{asympeq1}
 \lim_{m\to\infty} \frac{ Tr S_{m,k}(A,B)}{a_p^m {m+k \choose k}
 }=\tr C^k.
 \end{equation}
 \end{theo}

 We prove this theorem using the following lemmas.
 Recall that $B\in\rH_n$ is nonnegative definite if and only if its
 all principle minors are nonnegative.  Thus we obtain.

 \begin{lemma}\label{vanishrc}  Let $B=[b_{ij}]_{i,j=1}^n
 \in\rH_{n,+}$.  Then for each $i\ne j\in\an{n}$ $b_{ii}b_{jj}\ge
 |b_{ij}|^2$.  In particular, if $b_{ii}=0$ then $b_{ij}=b_{ji}=0$
 for $j=1,\ldots,n$.
 \end{lemma}

 \begin{lemma}\label{vantrace}  Let $A,B\in \rH_{n,+}$.  Then $AB=0$
 if and only if $\tr AB =0$.
 \end{lemma}

 \proof  Clearly, if $AB=0$ the $\tr AB=0$.  Assume that $\tr AB=0$.
 As above we may assume that $A$ is a diagonal matrix of the form
 (\ref{Adiag}) and $B=[b_{ij}]_{i,j=1}^n$.
 So $Tr(AB)=\sum_{i=1}^n a_i b_{ii}$.
 Since $B\succeq 0$ each
 $b_{ii}\ge 0$.  So $Tr(AB)=0$ yields that $a_ib_{ii}=0,i=1,\ldots,n$.
 Since $a_1\ge\ldots\ge a_n\ge 0$ we deduce from Lemma \ref{vanishrc}
 that $A=A_1\oplus 0_{t\times t}, B=0_{s\times s}\oplus B_2$, where
 $A_1\in \rH_{s,+}, B_2\in \rH_{t,+}$ and $s,t$ are nonnegative
 integers such that $s+t=n$.  Clearly $AB=0$.
 \qed

 \begin{lemma}\label{bidiagbm}  Let $A,B\in \C^{n\times n}$ and
 $2\le k\in\N$.
 Denote by $T_k(A,B)\in \C^{kn\times kn}$ the following $k\times k$
 block bidiagonal Toeplitz matrix
 $$T_k(A,B)=\left[\begin{array}{ccccccc}A&B&0&0&\dots&0&0\\0&A&B&0&\dots&
 0&0\\ \vdots&\vdots&\vdots&\vdots&\vdots&\vdots&\vdots\\0&0&0&0&\dots&A&B\\
 0&0&0&0&\dots&0&A\end{array}\right].$$

 Then for $m\in N$ $T_k(A,B)^m$ is the following upper triangular
 Toeplitz matrix:
 $\left[\begin{array}{cccccc}S_{m,0}(A,B)&S_{m-1,1}(A,B)&S_{m-2,2}(A,B)
 &\dots&S_{m-k+2,k-2}(A,B)&S_{m-k+1,k-1}(A,B)\\0&S_{m,0}(A,B)&S_{m-1,1}(A,B)&\dots&
 S_{m-k+3,k-3}(A,B)&S_{m-k+2,k-2}(A,B)\\ \vdots&\vdots&\vdots&\vdots&\vdots&\vdots
 \\0&0&0&\dots&S_{m,0}(A,B)&S_{m-1,1}(A,B)\\
 0&0&0&\dots&0&S_{m,0}(A,B)\end{array}\right]$.

 \end{lemma}

 \proof The lemma follows straightforward by induction from
 (\ref{basid}).

 Recall the Neumann expansion
 \begin{equation}\label{Neumexp}
 (I_{kn} -t T_k(A,B))^{-1}=\sum_{m=0}^{\infty} t^m
 T_k(A,B)^m.
 \end{equation}
 Let
 \begin{equation}\label{blockinv}
 (I_{kn} -t T_k(A,B))^{-1}=((I_{kn} -t
 T_k(A,B))^{(-1)}_{ij})_{i,j=1}^k.
 \end{equation}
 Then Lemma \ref{bidiagbm} and (\ref{Neumexp}) yield.

 \begin{eqnarray}\label{1basid}
 (I_{kn} -t T_k(A,B))^{(-1)}_{1k}=\sum_{m=0}^{\infty} t^m
 S_{m-k+1,k-1}(A,B)=\\
 t^{k-1}\sum_{m=k-1}^{\infty} t^{m-k+1}S_{m-k+1,k-1}(A,B).
 \nonumber
 \end{eqnarray}

 \begin{lemma}\label{2basid}  Let $A,B\in \C^{n\times n}$.
 Then
 \begin{equation}\label{2basid1}
 (I-t A)^{-1}(B(I-tA)^{-1})^{k-1}=\sum_{m=k-1}^{\infty}
 t^{m-k+1}S_{m-k+1,k-1}(A,B).
 \end{equation}

 \end{lemma}

 \proof  Observe that
 $$I_{kn}-tT_k(A,B)=T_{k}(I-tA,-tB)=T_k(I,-tB(I-tA)^{-1})T_k(I-tA,0).$$
 Hence
 \begin{eqnarray*}
 (I_{kn}-tT_k(A,B))^{-1}=T_k(I-tA,0)^{-1}\;T_k(I,-tB(I-tA)^{-1})^{-1}=\\
 T_k((I-tA)^{-1},0)\;T_k(I,-tB(I-tA)^{-1})^{-1}.
 \end{eqnarray*}
 Observe next that for any $C\in \C^{n\times n}$
 $T_k(I,-C)=I_{nk}-T_k(0,C)$, where $T_k(0,C)$ is nilpotent.
 Hence $T_k(I,-C)^{-1}=I_{kn}+\sum_{m=1}^{k-1} C^m$.  Furthermore
 the $(1,k)$ block of $T_k(I,-C)^{-1}$ is $C^{k-1}$.  Hence
 the $(1,k)$ block of $(I_{kn}-tT_k(A,B))^{-1}$ is equal to
 $t^{k-1} (I-t A)^{-1}(B(I-tA)^{-1})^{k-1}$.  Use (\ref{1basid}) to
 deduce (\ref{2basid1}).  \qed

 \begin{corol}\label{eqfrmbmv}  Let $A,B\in\rH_{n,+}$.  Then the BMV
 conjecture is equivalent to the statement that
 for each $k\ge 1$ the Taylor series of
 $\tr (I-t A)^{-1}(B(I-tA)^{-1})^{k-1}$ are nonnegative.
 \end{corol}

 \textbf{Proof of Theorem }\ref{fasymin}.
 The case $\tr AB=0$ is taken care by Lemma \ref{vantrace}.
 Assume that $\tr AB>0$.  We will prove first the equality
 (\ref{asympeq1}).  Assume that $A$ of the
 form (\ref{Adiag}).   Then
 $(I-tA)^{-1}=\diag((1-a_1t)^{-1},\ldots,(1-a_nt)^{-1})$.
 Let $B=[b_{ij}]_{i,j=1}^n$.  Then
 \begin{equation}\label{exfrmtr}
 \tr
 (I-tA)^{-1}(B(I-tA)^{-1})^{k}=\sum_{i_1=i_{k+1},\ldots,i_{k1}=1}^n \frac{\prod_{j=1}^{k}
 b_{i_ji_{j+1}}}{\prod_{j=1}^{k+1}(1-a_{i_j}t)}, \;
 \;k\ge 1.
 \end{equation}
 Assume first that $b_{11}>0$,  i.e. the value of $p$ in the
 statement of the theorem is $1$.
 So $a_1=\ldots=a_l>a_{l+1}$.
 Clearly the rational function given in (\ref{exfrmtr}) has poles at
 most at the points $z=\frac{1}{a_i}$ for all $i$ such that $a_i>0$.
 We now show that the $\frac{1}{a_1}$ is a pole of order $k+1$
 exactly.  Clearly, the coefficient of the term $(1-a_1t)^{k+1}$
 is obtained by letting $i_1=i_{k+1},\ldots,i_k$ range from $1$ to
 $l$.  Hence this coefficient is equal to $\tr C^k$, where
 $C=[b_{ij}]_{i,j=1}^l$.  Since $B\succeq 0$ it follows that
 $C\succeq 0$.  Assume that $\lambda_1(C)\ge \ldots \lambda_l(C)\ge
 0$.  The maximal characterization of $\lambda_1(C)$ yields that
 $\lambda_1(C)\ge b_{11}$.  Hence
 \begin{equation}\label{trckin}
 \tr C^k=\sum_{i=1}^l \lambda_j(C)^k\ge \lambda_1(C)^k\ge b_{pp}^k
 >0,
 \end{equation}
 where $p=1$.
 Write
 $$\tr (I-tA)(B(I-tA)^{-1})^{k}=\frac{\tr C^k}{(1-a_1t)^{k+1}}+f_k(t,A,B).$$
 Note that $f_k(t,A,B)$ may have poles only at $\frac{1}{a_i}$.
 Furthermore, the order of the pole at $\frac{1}{a_1}$ is at most
 $k$.  Hence asymptotically, the  contribution to the $m$
 coefficient of the power series of $\tr (I-tA)(B(I-tA)^{-1})^{k}$
 is from the term $\frac{\tr C^k}{(1-a_1t)^{k+1}}$.  This establish
 the equality (\ref{asympeq1}).  Use (\ref{trckin}) for $p=1$ to
 deduce (\ref{asympineq1}) from (\ref{asympeq1}).

 Suppose now that the value of $p$ in the theorem is greater than
 $1$.  From the arguments of the proof of Lemma \ref{vantrace}
 it follows that
 \begin{eqnarray*}
 A=A_1\oplus A_2,\; A_1=\diag(a_1,\ldots,a_{p-1}),\;
 A_2=\diag(a_i,\ldots,a_n),\\
 B=0_{(p-1)\times(p-1)}\oplus B_2,\; B_2=[b_{ij}]_{i,j=p}^n.
 \end{eqnarray*}
 Then $\tr (I-tA)(B(I-tA)^{-1})^{k}=\tr
 (I-tA_2)(B_2(I-tA_2)^{-1})^{k}$ and the theorem follows from the
 previous discussed case.  \qed

 \section{The case $n=k=3$}
 The first nontrivial case of the BMV conjecture is is the dimension $n=3$, i.e.
 $3\times 3$ matrices.
 The first nontrivial case $k=3$ is that we consider the
 coefficient of $t^3$ in $\tr (A+tB)^m$ for all $m$ and all $3\times 3$ matrices
 nonnegative definite matrices $A,B$.
 Assume for simplicity that we deal with real symmetric
 matrices.  Then the nontrivial case

 \begin{equation}\label{bform} B=\left[\begin{array}{rrr} x&-u
 &-v\\-u&y&-w\\-v&-w&z\end{array}\right], \quad u,v,w>0.
 \end{equation}
 It is enough to consider the case where $A=\diag(1,a,0)$,
 where $a\in [0,1]$ and $\det B=0$ and $B$ nonnegative definite.  This is equivalent to

 \begin{equation}\label{detb0}
 2uwv=xyz-u^2z-v^2y-w^2x,\;0\le x,y,z,\;u^2\le xy,\;v^2\le
 xz,\;w^2\le yz.
 \end{equation}
 Thus we need to show that all the coefficients in Taylor
 series of $\tr(I-tA)^{-1}(B(I-tA)^{-1})^3$ are all
 nonnegative.  Clearly

 $$B(I-tA)^{-1}=\left[\begin{array}{rrr}
 \frac{x}{1-t}&-\frac{u}{1-at}
 &-v\\-\frac{u}{1-t}&\frac{y}{1-at}&-w\\-\frac{v}{1-t}&-\frac{w}{1-at}&z\end{array}\right].
 $$
 Now compute the diagonal entries of $(B(I-tA)^{-1})^3$
 \begin{eqnarray*}
 ((B(I-tA)^{-1})^3)_{11}=\frac{x^3}{(1-t)^3}+\frac{xu^2}{(1-t)^2(1-at)}+\frac{xv^2}{(1-t)^2}
 +\\ \frac{yu^2}{(1-t)(1-at)^2}+\frac{zv^2}{1-t}-\frac{2u v w}{(1-t)(1-at)},
 \\
 ((B(I-tA)^{-1})^3)_{22}=\frac{y^3}{(1-at)^3}+\frac{yu^2}{(1-t)(1-at)^2}+\frac{yw^2}{(1-at)^2}+\\
 \frac{xu^2}{(1-t)^2(1-at)}+\frac{zw^2}{(1-at)^2}-\frac{2u v w}{(1-t)(1-at)},\\
 ((B(I-tA)^{-1})^3)_{33}=z^3+\frac{zv^2}{(1-t)}+\frac{zw^2}{(1-at)}+\\
 \frac{xv^2}{(1-t)^2}+\frac{yw^2}{(1-at)^2}-\frac{2u v
 w}{(1-t)(1-at)}.
 \end{eqnarray*}
 Hence $\tr(I-tA)^{-1}(B(I-tA)^{-1})^3$ is
 \begin{eqnarray*}
 \frac{1}{1-t}\big(\frac{x^3}{(1-t)^3}+\frac{xu^2}{(1-t)^2(1-at)}+\frac{xv^2}{(1-t)^2}
 +\\ \frac{yu^2}{(1-t)(1-at)^2}+\frac{zv^2}{1-t}-\frac{2u v
 w}{(1-t)(1-at)}\big)+
 \\ \frac{1}{1-at}
 \big(\frac{y^3}{(1-at)^3}+\frac{yu^2}{(1-t)(1-at)^2}+\frac{yw^2}{(1-at)^2}+\\
 \frac{xu^2}{(1-t)^2(1-at)}+\frac{zw^2}{(1-at)^2}-\frac{2u v w}{(1-t)(1-at)}\big)+\\
 z^3+\frac{zv^2}{(1-t)}+\frac{zw^2}{(1-at)}+\\
 \frac{xv^2}{(1-t)^2}+\frac{yw^2}{(1-at)^2}-\frac{2u v
 w}{(1-t)(1-at)}.
 \end{eqnarray*}

 So one has to show that all the coefficients of these series
 are nonnegative under the conditions (\ref{detb0}).

 \bibliographystyle{plain}

\begin{thebibliography}{MMM}
 \bibitem{BMV} D. Bessis, P. Moussa and M. Villani, Monotonic converging variational approximations
 to the functional integrals in quantum statistical mechanics,
 \emph{J. Mathematical Phys.}  16  (1975), no. 11, 2318--2325.
 \bibitem{Fle} C. Fleischhack, Asymptotic Positivity of Hurwitz Product
 Traces, arXiv:0804.3665.
 \bibitem{Hil} C.J. Hillar,
 Advances on the The Bessis-Moussa-Villani Trace Conjecture, Linear Algebra and Applications,
 to appear.
 \bibitem{LS} E.H. Lieb and R. Seiringer, Equivalent forms of the
 Bessis-Moussa-Villani conjecture, \emph{J. Statist. Phys.}  115  (2004), 185--190.

  \end{thebibliography}
 
 \end{document}